\def\pn{\par\noindent}
\def \SAIT #1 #2 {{\em Mem.\ Soc.\ Astron.\ It.\/} {\bf #1}, #2}
\def \MESS #1 #2 {{\em The Messenger\/} {\bf #1}, #2}
\def \ASTRNACH #1 #2 {{\em Astron. Nach.\/} {\bf #1}, #2}
\def \AAP #1 #2 {{\em Astron. Astrophys.\/} {\bf #1}, #2}
\def \AAL #1 #2 {{\em Astron. Astrophys. Lett.\/} {\bf #1}, L#2}
\def \AAR #1 #2 {{\em Astron. Astrophys. Rev.\/} {\bf #1}, #2}
\def \AAS #1 #2 {{\em Astron. Astrophys. Suppl. Ser.\/} {\bf #1}, #2}
\def \AJ #1 #2 {{\em Astron. J.\/} {\bf #1}, #2}
\def \ANNREV #1 #2 {{\em Ann. Rev. Astron. Astrophys.\/} {\bf #1}, #2}
\def \APJ #1 #2 {{\em Astrophys. J.\/} {\bf #1}, #2}
\def \APJL #1 #2 {{\em Astrophys. J. Lett.\/} {\bf #1}, L#2}
\def \APJS #1 #2 {{\em Astrophys. J. Suppl.\/} {\bf #1}, #2}
\def \APSS #1 #2 {{\em Astrophys. Space Sci.\/} {\bf #1}, #2}
\def \ASR #1 #2 {{\em Adv. Space Res.\/} {\bf #1}, #2}
\def \BAIC #1 #2 {{\em Bull. Astron. Inst. Czechosl.\/} {\bf #1}, #2}
\def \JSQRT #1 #2 {{\em J. Quant. Spectrosc. Radiat. Transfer\/} {\bf #1}, #2}
\def \MN #1 #2 {{\em Mon. Not. R. Astr. Soc.\/} {\bf #1}, #2}
\def \MEM #1 #2 {{\em Mem. R. Astr. Soc.\/} {\bf #1}, #2}
\def \PLR #1 #2 {{\em Phys. Lett. Rev.\/} {\bf #1}, #2}
\def \PASJ #1 #2 {{\em Publ. Astron. Soc. Japan\/} {\bf #1}, #2}
\def \PASP #1 #2 {{\em Publ. Astr. Soc. Pacific\/} {\bf #1}, #2}
\def \NAT #1 #2 {{\em Nature\/} {\bf #1}, #2}

\catcode`\@=11
\def\gsim{\ifmmode{\mathrel{\mathpalette\@versim>}}
    \else{$\mathrel{\mathpalette\@versim>}$}\fi}
\def\lsim{\ifmmode{\mathrel{\mathpalette\@versim<}}
    \else{$\mathrel{\mathpalette\@versim<}$}\fi}
\def\@versim#1#2{\lower 2.9truept \vbox{\baselineskip 0pt \lineskip
    0.5truept \ialign{$\m@th#1\hfil##\hfil$\crcr#2\crcr\sim\crcr}}}
\catcode`\@=12
\def\ref{\par\noindent\hangindent=1truecm}
\def \mg2{\mbox{Mg$_2$~}}
\def \mto{\mbox{$M_{\rm TO}$}}
\def \menv0{\mbox{$M_{\rm env}^{\rm HB}$}}
\def \msun{\mbox{$M_{\odot}$}}
\def \dydz{\mbox{$\Delta Y/\Delta Z$}}
\def \yr-1{\hbox{${\rm yr}^{-1}$}}
\def \lsun{\hbox{$L_\odot$}}
\def \zsun{\hbox{$Z_\odot$}}

\def \lt{\hbox{$L_{\rm T}$}}
\def \luv{\hbox{$L_{\rm UV}$}}
\def \te{\hbox{$T_{\rm eff}$}}
\def \hbte {\mbox{$T_{\rm eff}^{\rm HB}$}}

\documentstyle{memsait}
\input epsf.sty
\begin{opening}
\title{THE UV EMISSION OF ELLIPTICAL GALAXIES} 
\author{LAURA GREGGIO$^{1,2}$ and ALVIO RENZINI$^3$}
\institute{$^1$Universit\"ats Sternwarte M\"unchen, Scheinerstr. 1, D-81679
M\"unchen,  Germany\\
$^2$Dipartimento di Astronomia di Bologna, Via Zamboni 33, I-40126
Bologna, Italy\\
$^3$ESO, Karl-Schwarzschild-Str. 2, D-85748 Garching b. M\"unchen,  Germany}
\date{} 
\end{opening}

\begin{document}

\oddpagefooter{}{}{} 
\evenpagefooter{}{}{} 
\ 
\bigskip

\begin{abstract}
Observational as well as theoretical developments subsequent to 1990
concerning the origin of the ultraviolet emission of elliptical
galaxies are reviewed and discussed. In particular, successes and
failures of an extensive set of predictions contained in Greggio \&
Renzini (1990) are discussed in the light of these development.
Strong support to the preferred hot horizontal branch and AGB--manqu\'e
scenario for the origin of the UV emission has come
from UIT, HUT and HST observations. However, the UV sensitivity of
such space telescopes has proved insufficient to detect the 
predicted fade away effect of the UV upturn with redshift, that we
consider the ultimate check for this scenario. Some serendipitous
discoveries prompted by studies of the UV upturn are also mentioned.

\end{abstract}

\section{Introduction}

At optical wavelengths the spectral energy distribution (SED) of
elliptical galaxies falls precipitously shortward of the 4000 \AA \ break,
hence the discovery of the UV-rising branch came as one of the 
most unexpected results of the first UV satellites.  
The general prejudice was that ellipticals contained exclusively old and
cool stellar populations, similar to metal rich galactic globulars, 
albeit even more
metal rich. However, a few bright very hot stars were known to
be globular cluster members, and the presence of similar objects in
ellipticals had been suggested before the discovery of their
UV rising branch (Minkowski \& Osterbrock 1959; Hills
1971), but this knowledge was not widely spread.
The first UV spectra of ellipticals and bulges of spirals showed
instead that shortward of $\sim$ 2300 \AA~ the flux was increasing with
decreasing wavelenth. A fundamental contribution to this subject is
due to Burstein et al. (1988), who collected and organized all the relevant
information from IUE observations. The three main
results of this study were: 1) all studied
ellipticals have detectable UV flux, 2)  their (1550--V) 
color spans a range of $\approx$ 2.5 mag, and 3) it is 
strongly correlated with the \mg2 index. Hence, the ratio of the UV  
to the optical emission varies by $\sim$ an order of magnitude, 
and this ratio appears to increase with  average metallicity ($Z$),
assuming that the \mg2 index traces $Z$.

The presence of young, massive (hence hot) stars in ellipticals was
soon entertained in order 
to account for the observed UV radiation.
On the other hand, low mass stars do evolve through hot evolutionary
phases at the end of their life, and some UV radiation {\it is}
naturally expected to arise also from purely  old stellar populations.
Greggio \& Renzini (1990, hereinafter GR90) explored a variety of 
possible candidates
produced by the evolution of low mass stars, both single and in binary
systems. In particular, in GR90 we concentrated on the possibility of
producing hot stars in $\gsim$ 10 Gyr old stellar systems, which could
account for the observed level of the UV-to-optical flux ratio, and
(qualitatively) of the correlation with metallicity. 
We used simple energetic arguments, based on the fuel consumption
theorem (Renzini and Buzzoni 1986), to translate the observed
level of the UV rising branch
into specific requirements for the candidate stars
responsible for the UV emission. In this paper we first summarize the
main results of GR90, and then review and discuss both the
observational and the theoretical developments following 1990.

\section{The Theoretical Background}

The argument in GR90 goes as follows. 
The ultraviolet SED as measured from IUE is consistent with the
Rayleigh-Jeans tail of a black
body curve of temperature higher than $\sim$25000 K. In order to estimate  
the ratio of UV to total flux an assumption on the typical
temperature of the hot component is necessary. For example, for
NGC 4649, one of the most powerful ellipticals, $L_{\rm UV}/L_{\rm T}$
is in the range from 0.014 to 0.021, for $20000\lsim \te \lsim 40000$ K. 

For single age and single
metallicity populations of single stars (simple stellar populations, SSPs), 
the contribution to the total 
bolometric light of stars in the generic $j$-th post-MS evolutionary phase is 
\begin{equation}
\frac {L_{\rm j}}{L_{\rm T}} \simeq 9.75 \times 10^{10} B(t) F_{\rm j}(\mto)  
\label{eq:fct0}
\end{equation}
where $B(t)$ is the specific evolutionary
flux, i.e. the number of stars evolving through the turnoff and beyond
per year per solar luminosity of the parent population 
(in units of $\#**\ yr^{-1} \lsun^{-1}$), and $F_{\rm j}$ is the amount of
fuel burned during the 
phase $j$ by stars of initial mass equal to the turn-off mass (\mto) at
the age $t$ of the population. The fuel $F_{\rm j}$ is expressed in
\msun \ of equivalent hydrogen, i.e. $F_{\rm j} = \Delta M^{\rm
H}_{\rm j} + 0.1 \Delta M^{\rm He}_{\rm j}$, where
$\Delta M^{\rm H}_{\rm j}$ and $\Delta M^{\rm He}_{\rm j}$ are
respectively the mass of hydrogen and helium burned during the phase
$j$. For old SSPs $B(t) \simeq 2.2
\times 10^{-11}$ stars $L_\odot^{-1}$ yr$^{-1}$, almost independent of
composition and age (cf. Fig. 1 in Renzini 1998). 
Going one step further, equation (1)
can be generalized to a collection of SSPs (composite stellar
population, CSP), e.g. one exhibiting a narrow range of ages but a wide 
metallicity distribution, that GR90 adopted as a fair description 
of the stellar content of ellipticals.
To this end, suffice to substitute $F_{\rm
j}(\mto)$ with the fuel averaged over the metallicity distribution
$< F_{\rm j}>_{Z}$. Thus, for old stellar populations with a
metallicity distribution the following simple relation holds: 
\begin{equation}
L_{\rm j}/L_{\rm T}\simeq 2 \times <F_{\rm j}(\mto)>_Z.
\label{eq:fct}
\end{equation}
This equation was the main tool used in GR90 to evaluate various kinds
of stars as potential contributors to the UV rising branch in ellipticals.

From the observational requirement $L_{\rm UV}/L_{\rm T}\simeq
0.02$, equation (2) immediately indicates that
the hot stars responsible for the UV emission from giant elliptical
galaxies should burn at least 
$\sim$ 0.01 $M_\odot$ of equivalent hydrogen.
As already mentioned, the range in $(1550-V)$ colors spanned by ellipticals
is consistent with $L_{\rm UV}/L_{\rm T}$ varying by one
order of magnitude. Accordingly, the fuel burned by the candidate hot
stars, averaged over the metallicity distribution,
should increase from $\simeq$ 0.001 to $\simeq$ 0.01 \msun \ when the
average metallicity of the CSP increases. 
GR90 listed four candidates, which are naturally produced in the advanced
stages of the evolution of single stars:
\par\noindent
(i) Post AGB stars (P--AGB), i.e. stars which leave the AGB after the first
thermal pulse, and reach $\te\gsim 100,000$ K before approaching the
white dwarf (WD) cooling sequence. This is certainly the most common
channel to reach the finale WD stage. Typical luminosity $\sim 1000\lsun$.
\par\noindent
(ii) Post Early AGB stars (P--EAGB), i.e. stars leaving the AGB before
the first thermal pulse, as most of their hydrogen envelope is lost
before. Typical luminosity $\lsim 1000\lsun$.
\par\noindent
(iii) Hot HB stars (HHB), sometimes also called Extreme HB (EHB)
stars, i.e. stars which spend the core helium
burning phase at high temperatures, and whose subsequent evolution
(shell helium burning phase) also takes place at high temperature 
(AGB--manqu\'e). Typical luminosity $\sim 20\lsun$ for HHB, few
$100\lsun$ for AGB--manqu\'e.
\par\noindent
(iv) Post RGB stars (P--RGB), i.e. stars which fail helium ignition
because they loose their envelope while climbing on the RGB.
Typical luminosity $\lsim 1000\lsun$.

The first three channels eventually produce carbon-oxygen WDs, the
last one helium WDs. Fig. 1 shows schematically the evolutionary paths
corresponding to channels (i), (ii) and (iii). Also shown are the
limiting magnitudes for objects in M31 reached with 1.5 h exposures
with WFPC2 in two Wood's filters. It appears evident how difficult it
is to detect individual HHB stars at this distance.

A stellar population of given metallicity will
certainly produce stars evolving through channels ii), iii) and iv)
provided it becomes sufficiently old. However, the age at which this
happens cannot be accurately predicted.
A model star of given initial mass will evolve through
one of the four channels above depending on the wind mass loss rate 
efficiency ($\eta$). For $\eta$  below a
critical value it will go through the P--AGB, and for larger and larger
values of $\eta$ it will switch to the P--EAGB, HHB+AGB-manq\'e, and finally
to the P--RGB track. As illustrated in GR90, 
this whole range of possibilities is
realized by varying the mass loss rate parameter $\eta$ by just $\sim$
a factor of two, i.e. by an amount vastly smaller 
than any observational uncertainty affecting empirical RGB and AGB
mass loss rates. This leaves ample freedom to theoreticians.

\begin{figure}
\epsfysize=9cm 
\hspace{2.0cm}\epsfbox{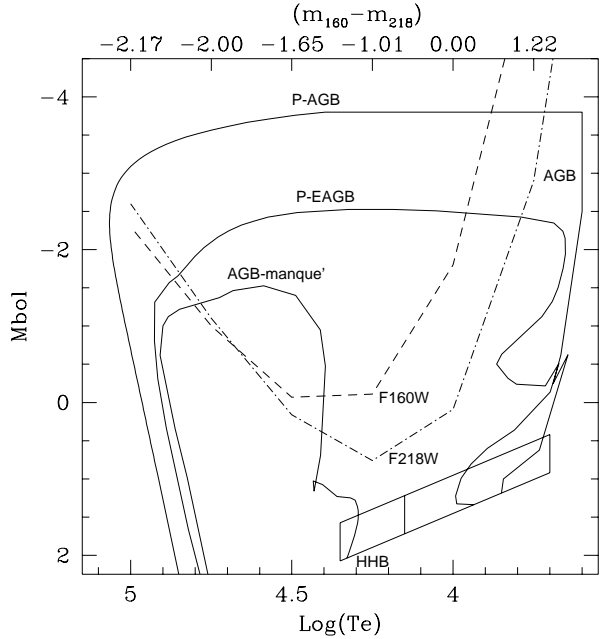} 
\caption[h]{Examples of evolutionary tracks for P--AGB, P--EAGB and
HHB plus AGB-manqu\'e objects. The slanted box indicates the HB locus.
Also shown are the limiting magnitudes
for objects at the distance of M31 (having adopted a true distance
modulus of 24.2) for 1.5 h exposures with WFPC2 (PC chip) on board of 
HST, in the two Wood's filter F160W and F218W.}
\end{figure}

All available evolutionary calculations indicated (and still indicate)
that P--AGB stars burn  less than $\sim 0.003$ \msun,
and P--RGB objects even less. This allowed GR90 to conclude that stars
in
these evolutionary phases could play only a minor role in the
production of the UV upturn.
More promising appeared the
P-EAGB, with $F_{\rm P-EAGB}$ up to $\simeq 0.025$ \msun, and the
HHB and their AGB--manqu\'e progeny,
burning in total $\sim$ 0.5 \msun \ of helium (equivalent to $\sim$ 0.05
\msun \ of hydrogen).
If all stars were to go through the HHB+AGB--manq\'e channel $\sim$ 5
times more UV radiation would be produced than
needed to account for the $\sim 2\%$ of the total luminosity emitted
in the UV, as in the $(1550-V)$ bluest galaxies. 
Thus, a relatively small fraction ($\approx$ 20 $\%$) of the 
stellar population in
ellipticals needed to evolve through channel (iii) in order to fit the
observations. 
The trend of the (1550--V) color increasing with $Z$ could
then be understood if a larger fraction of the population was evolving
through channel (iii) at higher average metallicity. 
Such a trend
could be accomplished in either of two ways: 1) with a modest increase
with metallicity of the mass loss rate parameter $\eta$, or 2) with the
helium abundance ($Y$) increasing with metallicity ($Z$). Indeed,
at fixed age and metallicity, a larger $Y$ corresponds to
a smaller envelope mass for the star evolving along the RGB, so that it is
easier to produce objects (ii) to (iv). Moreover, higher helium {\it
per se} favors higher effective temperatures, e.g. during the HB phase
(cf. Sweigart \& Gross 1976).
In essence, which
of the 4 channels is realized   depends on how the mass loss
rate and $Y$ scale with $Z$, both 
parameters $\eta(Z)$ and \dydz\ being poorly known  observationally. 
At the same time,
hosting stellar populations with a metallicity spread, ellipticals
should be inhabited by all four kind of objects, though in different
proportions.
  
The main conclusions in GR90 can be summarized as follows:
\par\noindent
(i) P--AGB stars, these hot low mass objects necessarily present in
ellipticals do not provide enough UV flux to account for the level of
the UV rising branch in the most powerful galaxies.
\par\noindent
(ii) The UV upturn in old stellar populations could be accounted for
only by the presence of  P--EAGB and HHB stars, with up to $\sim 20\%$
of all evolving stars venturing through these channels.
\par\noindent
(iii) The production of these stars at
high $Z$, as seemed implied by the $(1550-V)$ -- \mg2 correlation, was
possible within the uncertainties affecting the empirical
determination of  $\eta(Z)$ and \dydz, which are inputs to
stellar evolution theory. 
\par\noindent
(iv) Whatever mechanism is responsible for the production of these
stars (i.e. mass loss or \dydz), one expects that all possibile
candidates are present in ellipticals, though in different proportions. 
\par\noindent
(v) If the hot stars responsible for the UV emission in the most
powerful ellipticals were P--EAGB and HHB objects, the UV rising
branch should fade away rapidly with increasing redshift, possibly 
already disappearing at redshifts as low as 
$z \gsim 0.2$. This was a direct consequence of the sensitivity of
the effective temperature of HB stars on the envelope mass
at helium ignition.

GR90 paid less attention at the temperature distribution of the stars
evolving trough the various channels, but pointed out the
overwhelming difficulty of predicting such distribution, that from the
theoretical point of view depends on a number of arbitrary functions.
We felt indeed that detailed spectral synthesis modelling was not worth
the effort, while the only possible firm conclusions could be reached 
with very simple arguments.
However, it was mentioned that the $2300$ \AA \ dip in the SED requires
a gap in the temperature distribution between HHB stars and the
remaining, cool HB stars. No explanation  for the existence of this
gap was given in GR90. Moreover, if a continuity exists between HHB and P--RGB
stars, then one may expect HHB stars to extend all the way to the
helium main sequence, hence to fairly high effective temperatures
($\te \sim 50,000$ K). In other words, this scenario would most naturally
produce a fairly wide temperature distribution of UV emitters.
It was also speculated that a sizable fraction of HHB and their
AGB--manqu\'e progeny could be helium stars, suffice indeed fairly
modest mass loss rates ($10^{-10}-10^{-9}\msun\yr-1$) for these stars
to lose completely their hydrogen envelope. This possibility was
predicted to be subject to observational test, as in this case the UV
upturn should have exhibited some Wolf-Rayet, WN-like features, such
as low or absent CIV and Lyman lines, and strong HeII and NV.

\section{UV Observations Beyond IUE}
In 1990 much of the available information on the UV upturn came from
IUE, and it has been organized by Burstein et al. (1988). Later, most of the 
observational novelties came from UIT, HUT, and HST. Direct UV imaging
became possible, as well as spectroscopy down to the Lyman continuum.
  
\subsection{UIT and HST Imaging of Nearby Spheroids}

UIT and HST imaging have definitely ruled out massive stars as the
origin of the UV upturn in M32, the bulge of M31 , as well as in NGC~1399,
one of the UV most powerful ellitpicals (O'Connell et
al. 1992; King et al. 1992; Bertola et al. 1995a; Cole et al. 1998).

Extremely blue, low mass stars have been directly imaged in the
bulge of M31 and in M32 by HST (King et al. 1992; Bertola et al. 1995a; 
Brown et al. 1998). Although none of
these objects is a massive elliptical, they follow the $(1550-V)$ $-$ \mg2
correlation which characterizes all {\it quiescent} Es. 
The King et al. and Bertola et al. data were taken with the pre-Costar FOC. 
The first group obtained images of a central field in 
 M31 through a filter centered at $\lambda$ = 1750 \AA,
resolving more than 100 objects. Based on their measure of the UV
magnitudes, and on an upper limit to their $(1750 - B)$
color, the authours concluded that the resolved stars in the F175W images
are P--AGB stars; and, by comparing with the IUE flux from the same area,
that these stars 
account for only a fraction ($\sim 20 \%$) of the total flux at 1750 \AA.

Bertola et al. (1995a) obtained images of M31, M32 and 
NGC~205 through the combined UV filters F150W and F130LP, resolving
81, 10 and 78 stars in the three objects respectively. The point like
sources in NGC~205 were interpreted as young OB stars (as already
known from ground-based observations), while the
luminosity of the sources in M31 and M32 suggests that these are
P--AGB stars. By comparing with IUE data, the authors conclude that
the resolved P--AGB objects
can account for the total UV flux in the case of M32, while for M31 $\sim 50
\%$ of the UV flux comes from an unresolved background. 

Therefore, both groups conclude that the UV light in the bulge of M31
likely comes from the combination of P--AGB stars and 
fainter objects, which appear as an unresolved diffuse background on the HST
image. The different value derived by the two groups 
for the contribution of the resolved
sources results from the different assumptions on
the sensitivity calibration of the pre-COSTAR
FOC, and on the uncertainties on the red leak through the UV
filters. At any rate, the conclusions from both groups confirmed the
prediction of GR90, i.e. the population of hot stars in old
stellar systems is composite, with contribution from P--AGB stars
bright enough to be individually detected in nearby spheroids, and
fainter sources such as P--EAGB and HHB+AGB--manqu\'e stars as faint as 
$\sim 20\lsun$ (hence below detection threshold).
Concerning M32, the low level of its UV upturn is in agreement with
the notion that the UV sources in
this galaxy should just be P--AGB stars, with very few stars -- if any
-- going through the (ii)-(iii) channels.

UV color gradients have also been detected in a few objects 
(O Connell et al. 1992). With the exception of M32, UV
colors become redder with increasing radius, probably tracing the \mg2
gradients. 
The UV light appears diffuse, but more concentrated than
the visual light, in agreement with the expectation that it is
produced by the higher $Z$ stars, preferentially found in
the central regions (see also Brown et al. 1998).

Post-Costar FOC photometry of M32 and M31 has been recently
obtained by Brown et al. (1998) in two UV filters, namely F275W and
F175W. Again, many point like sources are resolved in these images:
433 stars in M31 and 138 in M32 down to $m_{\rm F275W} = 25.5$ mag and 
$m_{\rm F175W} = 24.5$ mag. Brown et al. (1998) show that the pre-COSTAR
FOC calibrations were likely in severe error, basically leading to an
overestimate of the intrinsic UV flux from the sources. As a result, the
resolved stars in Brown et al. are interpreted as AGB--manqu\'e objects, the
bright progeny of HHB stars. Again, the cumulative flux 
from the resolved stars accounts for only a
fraction ($< 20 \%$) of the total IUE flux.
Although still affected by some uncertainty, in particular a possible
systematic underestimate of the flux in the F275W filter at the 0.3
mag level, the photometry by Brown et al. (1998)  
is in reasonable agreement with the expectations from IUE and HUT
spectra. 

The interpretation of the nature of the resolved sources in Brown et
al. (1998) rests essentially upon the characteristic of the luminosity
functions (LF) in the two UV filters. There appears to be an increasing
number of objects towards fainter magnitudes, a trend which is not
present in the P--AGB tracks of Vassiliadis \& Wood (1994) to which
the empirical LF was compared. These tracks peak instead at magnitudes
for which there are virtually no stars observed at all. Brown et al. conclude
that the bulk ($\gsim 95\%$) of all stars do indeed go through the
P--AGB channel, but the mass of the P--AGB stars is in excess of
$0.63\msun$, for which the P--AGB timescale is so short to be
consistent with the observed LF. However, such a high value of the
P--AGB mass would imply the existence of a prominent and very bright
population of AGB stars, for which there is no evidence in the bulge
of M31 (Renzini 1998). Moreover, this population would produce an 
enourmous amount of
energy ($\gsim 0.15\msun$ of fuel would be burned on the AGB), hence
leading to optical--infrared colors at variance with the observed ones.
In our opinion, the Brown et al. LF demonstrates that the Vassiliadis
\& Wood P--AGB tracks are inapplicable to the case of the M31 bulge.
These tracks are based on the assumption that the transition from the 
AGB to the planetary
nebula stage takes place on a nuclear time scale, being
controlled by the burning of the residual envelope mass. For the low
values of the P--AGB mass expected in an old stellar population 
($\sim 0.55\msun$) this transition time is indeed very long ($\sim
10^5$
yr), and a sizable number of hot P--AGB stars would have been
observed,
lying along the nearly horizontal track in the upper part of Fig. 1.
However, one knows from galactic globular clusters that the transition is
instead much faster, taking place either on a mass-loss time scale, or
even more probably on a thermal time scale (Kaeufl, Renzini, \&
Stanghellini 1993, and references therein). We conclude that the
observed LF is likely due to the combination of the low-mass P--AGB 
channel ($\gsim 95\%$ of the total stellar evolutionary flux), 
with the transition to high temperatures taking
place on a thermal time scale, plus P--EAGB and/or the
HHB+AGB--manqu\'e objects for the residual stellar evolutioanry flux   
($\lsim 5\%$ of the total).

An apparently  puzzling result of the Brown et
al. study is that the LFs of
the UV stars in M31 and M32 look similar in shape, in spite of the
strong difference in the level of the UV upturn in these two galaxies. 
According to Brown et
al. the fraction of the total evolutionary flux that has to go through
the non P--AGB channel is 2 $\%$ and 0.5 $\%$ respectively in the
bulge of M31 and in M32. Hence, the similarity of the two LFs comes
from both being dominated by P--AGB stars that do {\it not} evolve on 
a nuclear time scale through their transition from the AGB to high 
temperatures.

In 1993 FOC imaging in four UV bands was obtained of the central
regions of the ellipticals NGC 1399 and NGC 4552 and of the bulge of
the NGC 2681 spiral (PI F. Bertola). The aim was to study the spacial 
structure of the UV emission, checking for color gradients and if any 
patcheness existed due to star formation. No such patcheness was
found neither strong color gradients, but instead NGC 4552 and NGC
2681 showed a central, unresolved, point-like source (Bertola et al. 1995b).
To our surprize, we found that the point-like source in NGC 4552 had 
changed its
brightness by a factor $\sim 7\pm 1.5$ in the F342W band, compared to
a previous FOC image taken in 1991: a central {\it flare} had been
discovered, possibly due to a red giant having been tidally stripped
by a massive central black hole (Renzini et al. 1995). Subsequent,
post-COSTAR FOC imaging and FOS spectroscopy confirmed that the
central source is a variable mini-AGN, possibly the faintest known
AGN, with broad (FWHM$\simeq 3000$ km s$^{-1}$) emission lines
(Cappellari et al. 1998). While trying to better understand the UV
upturn, we had serendipitously found yet another way of gathering
information on the central black hole demography in galaxies.

\subsection{HUT Spectroscopy of the UV Upturn}

Extending the observed spectral range down to the Lyman limit
HUT has detected 
the maximum in the UV spectral energy distribution (Ferguson et
al. 1991). To date, HUT data for 8 early type objects, including the
bulge of M31, have been collected (Ferguson $\&$ Davidsen 1993; Brown,
Ferguson $\&$ Davidsen 1995; Brown et al. 1997). In all studied objects, 
the UV rising branch appears to have a turn-over at
$\lambda \approx$ 1000 \AA, which indicates that the bulk of the
radiation comes from moderately hot stars, with temperatures in the range
20000--25000 K (Brown et al. 1995). 
Assuming that this is the characteristic spectral energy distribution
in the UV for giant ellipticals, like NGC~4649,
we obtain a better estimate for the ratio $L_{\rm UV}/L_{\rm T}$ of $\simeq
0.015$, which translates into $< F_{\rm j}>_{Z} \simeq 0.007$ for
the hot stars inhabiting the most powerful ellipticals. 

Since the UV SED has a minimum  
around 2300 \AA , a large contribution from stars with
intermediate temperatures, say $\approx$ 10000 K, is excluded. 
Thus the bulk of
the UV emission comes from stars in a narrow range of temperatures.
This is an important constraint for the astrophysically plausible
evolutionary paths that can account for the UV rising branch
phenomenon. For 
example, an even distribution of stars on the HB like in the globular
cluster M3, corresponds to a spectrum flatter than observed in ellipticals,
due to the similar contribution from stars in the wide effective
temperature range (see e.g. Nesci $\&$ Perola 1985, Ferguson 1995).

Another important characteristic of the HUT spectra of early type
systems is the fact that they are composite:
when removing from the observed spectrum the theoretical contribution
of HHB star, according to their complete evolution from the ZAHB to the
WD final stage, some residual flux at the shorter wavelenghts
is left (Ferguson and Davidsen 1993; Brown et
al. 1997). The best fits to the SED of all the studied objects are
obtained with contributions from both HHB and P--AGB evolutionary tracks. 
Based on their detailed modelling, Brown et al. (1997)
conclude that approximately 10 $\%$ of the total stellar population
should go through the HHB channel of evolution in NGC~1399, one of
the strongest UV emitters. This is in very good agreement with the
predictions of the fuel consumption theorem: 
for a two components CSP, with 90$\%$ of the stars
evolving through the P--AGB channel and the remaining 10$\%$ going
through the HHB evolution, the average fuel burned in the hot
evolutionary phases is:
\begin{equation}
< F_{\rm j}> = 0.9 \times F_{\rm P-AGB} + 0.1 \times F_{\rm H-HB}
\label{eq:fuelave}
\end{equation}
Adopting $F_{\rm P-AGB}$ = 0.003 \msun \ and $F_{\rm H-HB}$ = 0.05
\msun \ (see Sect. 2) one gets $< F_{\rm j}>$ = 0.0077, close
indeed to the estimate above.
 
All of the 8 objects in the Brown et al.\ sample seem to require some 
contribution from HHB stars, in
different proportions. This does not come unexpected, since their \mg2
indices range from 0.31 to 0.36, which puts them among the high average 
metallicity objects.
Brown et al. (1995) also claimed that, within their
sample, a larger fraction of
stars evolve through the HHB channel in the stronger \mg2 galaxies.
Modelling the UV spectra, they derive the stellar
evolutionary flux through the HHB plus AGB--manqu\'e track which is 
needed to account for the observed UV emission. The ratio
beteween this evolutionary flux and that of the total stellar
population sampled by the HUT aperture appears to be nicely correlated
with the \mg2 index of the parent galaxy.
The value of this ratio is model dependent, and somewhat
different figures are obtained in the more detailed computations
in Brown et al. (1997). Nevertheless, judging from their tables, 
the general trend seems confirmed. Thus, it can be concluded  
that galaxies with \mg2 indices in excess of $\sim$ 0.3 very likely 
host HHB stars in their nuclei. Only a small fraction of the
population need to go through this extreme evolutionary channel to
account for the observed UV fluxes, varying from
$\sim 1$ to $\sim 10 \%$ for galaxies with \mg2 ranging from 0.3 to
0.36. These results are clearly consistent with the expectations from
GR90.

The low S/N in the HUT spectra prevents accurate determinations of
abundances. 
The absorption features seem however to indicate a low
metallicity in the stellar atmospheres of the stars mainly
contributing the UV emission: $Z_{\rm atm} = 0.1 Z_\odot$ (Brown et al 1997).
This would imply that the UV rising branch phenomenon is not directly related
to the presence of high $Z$ stars, and the correlation between the
$(1550-V)$ color and \mg2 index has to find a different  
explanation (Park \& Lee 1997, see below) from what proposed in GR90. 
Alternatively, $Z_{\rm atm}$
is not representative of the true metallicity of the hot stars in
ellipticals, as heavy elements may diffuse out of the
atmospheres of HHB stars (Brown et al. 1997).
A more firm result of the abundance analysis is the lack of CIV as
would be expected if massive stars were responsible for the UV
emission. However, HUT spectra show also strong Ly$\beta$ and Ly$\gamma$
lines, indicating that the vast majority of HHB and AGB--manqu\'e do
{\it not} lose their hydrogen envelope, and do not become WN-like
helium stars. This means that average mass loss
rates during these phases must be lower than $\sim 10^{-10}$ and $\sim
10^{-9}\msun\yr-1$, respectively.

\subsection{Attempts at Detecting the Evolution with redshift of the 
UV Upturn}

One crucial prediction of GR90 concerned the evolution with redshift
of the UV upturn. If due to a combination of P-EAGB and
HHB+AGB--manqu\'e stars, the UV upturn should fade away already
at fairly low redshift, see for example the realization by 
Barbaro, Bertola, \& Burstein (1992).
To check this prediction two Cycle-I HST
projects were implemented (PIs R. Windhorst and A. Renzini, respectively).
FOS spectra of $z=0.1-0.6$ elliptical galaxies selected for being
either weak radiogalaxies and/or cluster members were obtained. They
all showed a strong UV upturn, which at first sight appeared to be in clear
conflict with the prediction. However, it soon turned out that a
similarly strong UV upturn was also shown by the FOS spectrum of
an innocent G2V star, which certainly did not have it on its own. 
While searching for the vanishing UV upturn of ellipticals the
red-scatterd light problem of FOS was discovered instead (Windhorst et
al. 1994). This lead to a novel approach to the calibration of FOS --
and lately of ESO instruments for the VLT -- which makes more use of
first physical principles, and less recourse to least square fits
(Bushouse, Rosa, \& M\"uller 1995; Rosa 1997;
Ballester \& Rosa 1997). 

\begin{figure}
\epsfysize=9cm 
\hspace{2.0cm}\epsfbox{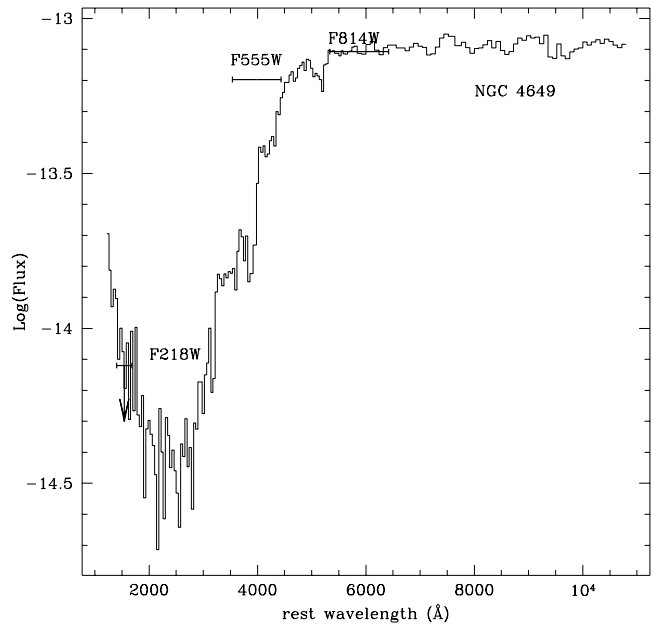} 
\caption[h]{The flux sampled by the F555W and F814W filters, and the 3
$\sigma$ upper limit to the flux through the F218W filter for the
brightest elliptical galaxy in the WFPC2 field of view in the $z=0.37$
cluster A895, as a function of the restframe wavelength. The SED of
NGC~4649 is also shown. The three
measured fluxes have been de-redshifted and normalized in such a way
to have the F814W flux to match the continuum of NGC~4649.}
\end{figure}

The FOS scattered light problem had the effect of reducing
dramatically the S/N ratio for UV observations of high-$z$
ellipticals, and therefore attempts at detecting the vanishing upturn
effect moved to WFPC2, now equipped with Wood's filters. 
A first attempt was made by a group including R. Gilmozzi, E. Held, R. Viezzer
and ourselves. WFPC2 images  of the cluster Abell 895 ($z=0.37$) were
obtained through the Wood's filter F218W, and through the 
F555W and F814W filters.
No detectable flux from cluster ellipticals was found in a coadded
10,000 second integration through the F218W filter.
The result is shown in Fig. 2 for the brightest cluster member
(reproduced from Renzini 1996), with the 3-$\sigma$ upper limit
falling disappointingly on top of the expected upturn if such
galaxies had the same rest frame $(1550-V)$ color of the local
elliptical NGC 4649.

Similarly disappointing was the result of an analogous experiment by 
Buson et al. (1998), who
imaged the Abell 851 cluster ($z=0.41)$ through the F218W 
 and F702W filters (corresponding to $\sim$ the rest
frame $(1550-V)$ color). Again, the F218W data are not deep enough to
detect the cluster ellipticals even if they were to maintain the same
rest frame $(1550-V)$ color of the bluest ellipticals at zero redshift.

The failures of these attempts has to be ascribed to the low sensitivity
of WFPC2 when used in conjunction with Wood's filters (that indeed we
nicknamed wood's filters). An alternative approach has been recently
pursued by  Brown et al. (1998) for a sample of ellipticals in the 
$z=0.375$ cluster
A370. The combination of two long-pass filters of FOC (F130LP and F370LP)  
has allowed them to isolate the contribution of the emission shortward of
$\sim$ 2700 \AA\ in the rest frame, hence sampling the UV upturn.
Surprizingly, no appreciable evolution compared to nearby ellipticals has been 
detected, and Brown et al. conclude that this result excludes some models 
of the upturn, while others are still acceptable provided that 
the bulk of stars
in these galaxies formed at $z\gsim 4$.
 
More observations are needed to study in detail the evolution of the
UV rising branch with increasing redshift, and to derive informations
on the nature and the age of the UV bright stars. Since FOS and GHRS have been
removed from HST, STIS may now offer a better chance to detect
the vanishing UV upturn  effect.

\section{Theoretical Modelling}

\subsection{Stellar Evolutionary Sequences}

In 1990 only a handful of P--AGB, P--EAGB, and HHB+AGB--manqu\'e
evolutionary sequences existed in the literature.
In the last decade a large effort has been devoted to
construct extensive sets of evolutionary tracks, primarily 
with the aim of understanding the UV upturn
phenomenon. Hundreds of stellar evolutionary 
sequences for low mass stars, with up to super solar metallicities and
helium abundances have been computed to isolate the range of
parameteres which produce P--EAGB and HHB objects
(e.g. Castellani \& Tornamb\'e 1991; Horch, Demarque \& Pinsonneault 1992; 
Castellani, Limongi \& Tornamb\'e 1992, 1995; Dorman, Rood and O' Connell
1993, hereinafter DRO93; Fagotto et al. 1994a,b,c; Yi, Demarque \& Kim 1997a).
Basically, the overall evolutionary picture
illustrated in GR90 has been confirmed. The average temperature at which the
helium burning occurs is essentially controlled by the envelope mass of the
stars at helium ignition, being
hotter the lower the envelope mass. High values of the helium
abundance favor the
production of hot helium burners, and widen the range of envelope
masses for which this condition is satisfied.
According to DRO93, for
$Z$= \zsun\, stellar models with HB envelope masses \menv0\ 
$\lsim$ 0.05 \msun\ evolve either as P--EAGB or as HHB
and AGB--manqu\'e. This critical value for
the envelope mass increases with the helium abundance 
(see also Yi et al. 1997a), reaching values as high as 0.15 
\msun\ for $(Y,Z)$ = (0.45,0.06). It follows that at high $(Y,Z)$ the 
condition on
the envelope mass necessary to produce hot stars is more easily met. 

Two interesting aspects of the evolution of low mass stars have been
disclosed, which were not considered in GR90: 1) for high $Z$ and
especially $Y$, 
the evolution of HB stars presents a pronounced dichotomy, with some
stars starting the evolution on the red side of the HB, spending there
a fraction of their HB phase, and then after 
zipping to high temperatures where they burn the rest of their fuel 
(Horch et al. 1992, but see also Sweigart \&
Gross 1976); and
2) for high assumed mass loss rates some
stars {\it peel off} the RGB and experience their core helium flash
at high effective temperatures (Castellani \& Castellani 1993; 
D'Cruz et al. 1996).

The systematics with $Y$ and $Z$ of the post RGB evolution can be 
appreciated in DRO93 and Yi et al. (1997a): up to $\approx$ \zsun
the dependence of the ZAHB \te\ on \menv0\ is
relatively mild, and a flat distribution of envelope masses maps into an
even distribution in Log $T_{\rm eff}$ of stars on the HB.
Subsequent evolution remains confined in the red (in the blue)
for the more massive (less massive) HB objects, while the intermediate
HB (IHB) stars evolve along wide redward/blueward loops, thereby
providing intermediate temperature objects. Correspondingly,
for $Z \lsim$ \zsun it is relatively difficult to produce the 
2300 \AA\ minimum in the SED. However, as the metallicity
increases, the evolution of the HB objects tends to become more
skewed either towards the red, or to the blue. The effect is very
strong for large values of the \dydz\ parameter. At ($Y,Z$)= (0.46,0.06) 
the IHB objects virtually disappear, and the bulk of stars are
either redder than Log $T_{\rm eff} \simeq$ 3.7 or bluer than Log
$T_{\rm eff} \simeq$ 4.2 (see Figures 2 and 3 in DRO93). This
behavior may help producing the 2300 \AA\ minimum in the SED. 

As illustrated in the introduction, when the mass loss parameter 
$\eta$ is sufficiently large, the evolution
on the RGB is aborted before the core mass has grown enough to
trigger the central helium flash. In GR90 it was assumed that in this
case further evolution would just take the model star to the (helium)
white dwarf stage. Actual computations in this $\eta$
range show, instead, that there are
models which succeed in igniting helium after departing from the RGB,
either while crossing the HRD, or during the subsequent cooling phase
towards the WD stage (Castellani \& Castellani 1993). 
Thus, there is a mass range (or a $\eta$ range) for which
the helium core flash occurs in the hot region of the HRD (hot helium
flashers, in D'Cruz et al. 1996 nomenclature).
Subsequent evolution of these objects (hereinafter HHeF) is the same
as for HHB star, with a very low envelope mass. If they exist, the HHeF 
have the minimum envelope mass that HB stars can have, hence
naturally defining the hot end of the horizontal branch. Stars with more
massive envelopes will ignite helium at the tip of the RGB, 
to appear on the ZAHB with lower \te. Stars less
massive than the HHeF will fail helium ignition, thus becoming helium WDs. 
At super solar metallicites, HHeF are produced 
for $\eta \gsim 0.7$, which is $\sim$ 2 times larger
than the value which fits the properties of the HB in
globular clusters (D'Cruz et al. 1996). After helium ignition, 
the HHeF are found on the
ZAHB at Log $T_{\rm eff}$ $\sim$ 4.4 for supersolar $Z$, which seems to
be the maximum possible temperature for HB models (Castellani,
Degl'Innocenti and Pulone 1995). 

\subsection{Synthetic UV Upturns}

Inspired by the $(1550-V)$ -- \mg2 correlation most authors have
explored under which conditions HHB and related stars are produced in
metal rich and super metal rich populations.

Dorman, O'Connell and Rood (1995, hereinafter DOR95) assume
the mass loss on the RGB as principal actor in
originating the UV rising branch. Two are the main points which
support this picture: (1) the presence of 
extended HB in the CMD of globular clusters, which require a spread in
RGB mass loss by $\sim$ 3 0$\%$ among stars within the same cluster; and (2)
the population of hot subdwarfs in the solar vicinity, which shows
that at $\sim$ solar metallicity HHB stars and their progeny can
occasionally be produced.  
Thus, in DOR95 view, the hot stars in ellipticals are
(moderately) old, $\sim$ solar
metallicity objects which happen to loose 2-3 times more mass than the
average Reimers rate. Questioning the real
significance of the \mg2 index as a metallicity indicator,
DOR95 generically ascribe the origin of the correlation 
$(1550-V)$ -- \mg2  to either an age or a metallicity
spread among ellipticals.
However, no attempt is made to eplore the effect of
a metallicity distribution on the UV SED of
ellipticals. If a large dispersion of the mass loss rate applies to
all $Z$ components, it seems difficult to avoid a sizeable contribution of
IHB stars in the UV spectral range.

In Tantalo et al. (1996) models HHB stars are produced at high $Z$
basically because a large \dydz\ is assumed. These authors construct 
self consistent
chemo-spectro-photometric models for the ellipticals, which thus
contain a metallicity distribution as computed from the chemical
evolution. The final integrated properties of the models depend not
only on the assumptions on the parameters governing the stellar
evolution (e.g. the mass loss), but also on those important for the chemical
evolution (i.e. star formation rate, IMF, stellar yield, depth of the
galactic potential, supernova feedback, galactic winds etc.). 
Tantalo et al. most massive galaxy models present 
a strong UV upturn developing as early as 5.6 Gyr. 
This value of the age is extremely sensitive to the specific choice of
the parameters $\eta$ = 0.45 and \dydz = 2.5, which cause the 
SSP model at $Z = 0.1$ to produce HHB stars already at 5.6 Gyr. 
On the other hand, all other SSP models in Tantalo et al. grid 
(with $Z < 0.1$) produce HHB stars only for ages in excess of $\sim$ 12 Gyr.
Therefore, the
UV properties of the composite model also depend critically on the precise
population of the highest metallicity bin (see also Yi, Demarque 
\& Oemler 1998). 

Finally, Yi, Demarque \& Oemler (1997b,1998) propose a model in which  
all the relavant parameters are allowed to vary while searching for a
best fit. They finally favour  
a positive (but moderate) \dydz (=2-3), a modest trend of $\eta$ with
metallicity (ranging from $\sim$ 0.5 to $\sim$ 0.7--1 for $Z$ ranging
from 0.02 \zsun to $\gsim$ \zsun) ; a mass dispersion of the HB of
$\sim$ 0.06 \msun (calibrated on GCs
properties); and a metallicity distribution, as suggested by chemical
evolution models. With these prescriptions, Yi et al. (1998) reach a
reasonable fit with the observations, at ages $\gsim$ 10 Gyr.
 The fit is better when adopting the $Z$ distribution
from infall models, as opposed to closed box models. Indeed, in the
latter case too much flux is produced in the mid-UV, due to the broad
distribution of HB temperatures of metal poor stars.  

While in all the above attempts the UV emission arises from stars in
the high-$Z$ tail of the metallicity distribution, Lee (1994) and Park
and Lee (1997) maintain 
that the UV flux
originates from the emission of metal poor stars. Considering that the
stellar populations in galaxies are characterized by a metallicity
distribution, they explore the possibility that the optical
light comes from the high $Z$ and the UV light come from the 
low $Z$ components.
The relatively low strength of absorption features in
the UV-rising branch mentioned in Section 3.2 is in agreement
with this picture.
The trend of increasing $(UV-V)$ color with increasing
\mg2 (hence $< Z >$) would have an indirect origin, resulting from 
the brighter ellipticals 
(with stronger \mg2 indices) being older than the fainter ones, as it
may be expected in some cosmological simulations.

However, in the Park and Lee model, ages as old as of $\sim$20 Gyr are
needed to produce a UV output such as that of giant ellipticals, which looks
uncomfortably too old. Another problem comes from
the strongest UV upturns being shown by  galaxies
with \mg2$\gsim$ 0.3, which cannot have a major metal poor component,
as the metallicity distribution shoud be trimmed below
Z$\approx$ 0.5 Z$_{\odot}$ (Greggio 1997).
Thus the low $Z$ tail may not be
present in the central regions of the most powerful ellipticals
(see also Chiosi, Vallenari \& Bressan 1997; Yi et al. 1998).
Moreover, as already noticed, it is more difficult to low-$Z$
populations to produce the 2300 \AA\ minimum in the SED, as the $T_{\rm
eff}$ of HB stars is a mild function of the envelope mass. Indeed,
the $(1500-2500)$ color of
the UV rising branch in ellipticals is systematically bluer than in
metal poor globular clusters (DOR95). This shows that the average
temperature of the hot stars in ellipticals is higher than the average
temperature of HB stars in GCs. Only high ($Y,Z$) models seem
to reach blue enough colors (DOR95, Yi et al. 1995, 1998).
Finally, we notice that two metal rich globular clusters in the
Galactic bulge have been found to host a sizable population of HHB
stars (Rich et al. 1997). This has shown that it is indeed possible to
produce numerous HHB stars at $Z \sim$ \zsun, although this 
may be due to some yet unidentified dynamical process in these
particular clusters.

To summarize, most of the theoretical work has just shown
quantitatively, in a detailed level, that the options are
equivalent: hot stars are produced at high metallicities in old
stellar populations if \dydz\ is large and/or if the mass loss rate
moderately increases with metallicity.
Which combination of the two effects is at work remains unclear. As we will
discuss later, high $Y$ in combination with high $Z$ could
be a necessary ingredient in order to avoid too much light in the
mid-UV from IHB stars.

\pn
\section {Discussion and Conclusions}

In summary, the current observational evidences that need to be
explained include:
\pn
\begin{itemize}
\item In those ellipticals with the strongest UV upturns
$L_{\rm UV} /L_{\rm T} \simeq$ 0.015. 
\pn 
\item The UV SED requires a small temperature range
for the hot stars in the nuclei of ellipticals, peaking at $\sim
25,000$ K. 
\pn
\item  Among ellipticals with \mg2 $\gsim$ 0.3, the fraction
of hot evolved objects going throuh the HHB channel of
evolution varies sizeably, possibly ranging from $\sim 1\%$ to $\sim 10\%$  
for \mg2 increasing from 0.3 to 0.36. 
\pn
\item The shape of the LF of the UV bright stars is similar in M32 and
in the bulge of M31. In both galaxies resolved stars account for only
of a small fraction of the UV flux.
\end{itemize}
\par
In this section we discuss the hints on the hot stars in
ellipticals from these observational evidences.

\subsection {On the hot stars in ellipticals and their origin}

Thanks to the HUT spectra reaching shorter wavelengths compared to
IUE, a smaller $\luv/\lt$ is derived, hence a smaller average fuel consumption
$< F_{\rm UV}>_{Z}$ is required
for the hot stars in the most powerful ellipticals, compared 
to GR90 estimates. However the average fuel consumption remains much
larger than what provided by P--AGB stars, and the best candidates hot
stars remain P--EAGB and/or HHB+AGB--manqu\'e objects.
HHB stars and their progeny tend to be favored essentially because
their $F_{\rm UV}$ is larger than the required $< F_{\rm j}>_{Z}$,
 so only a (small) fraction of the
stellar evolutionary flux has to go through a very hot helium burning
stage. Besides, P--EAGB stars are likely to be distributed over the
whole \te\  range from $\sim$ 5000 to $\sim$ 70000 K (see tracks in 
Castellani and Tornamb\'e 1991), thus providing too much flux both in the
mid-UV and in the most extreme UV spectral range.

If the bulk of hot stars in giant Es are HHB + AGB--manqu\'e, 
some $\sim 10 \%$ of the evolving population in the
UV brightest ellipticals has to evolve through channel (iii) (see
Section 2). This
constrains the combination of the parameters $\eta$ and \dydz, plus
all the parameters which play a role in determining the metallicity
distribution in the central regions of the most powerful ellipticals,
in particular its high metallicity tail. 

Can we learn something on these parameters 
from the UV SED?
As repeatedly noticed here, the observations require that the HHB
stars are characterized by a narrow range of \menv0. In this respect,
stellar populations of high metallicities, possibly coupled with large
helium abundances, are favored. This stems from an observational
argument, since the SEDs of galactic (low $Z$) globulars with extended HBs 
tend to be flatter than those typical of giant ellipticals.
At the same $(1500-V)$ color (i.e. for the same average $F_{\rm UV}$),
the $(1500 - 2500)$ color of GCs are redder than those of giant
ellipticals, with only high $Z$ SSP models matching the $(1500
- 2500)$ color (DOR95, Yi et al. 1995, 1998). 
There is also a theoretical argument in favor of
the high $Z$ hypothesis, based on the shape of the relation between
\hbte\ and \menv0\ of HB stars. As \menv0\ decreases \hbte\ keeps low untill a
threshold value is reached, after which the relation becomes
becomes extremely steep (cf. Fig. 3.1 in Renzini 1977). 
As a consequence there is 
a very narrow range of \menv0 which corresponds to intermediate
effective temperatures.
Perhaps more importantly, at high metallicites HB stars appears to exhibit a
bimodal behavior, with most of the HB (as well as the subsequent
shell helium burning stage) being spent either at high or low \te,
virtually avoiding the intermediate regime (see tracks in DRO93). This
tendency is reinforced when high $Y$ combines 
with high $Z$ (Yi et al. 1997b). 

In conclusion, the presence of the 2300 \AA\ minimum and the relatively 
steep slope of the UV SED in ellipticals speak in favor of high $Z$ and $Y$
HHB stars and their progeny.
At this point we notice that the strong \mg2 index in the nuclei of
giant Es requires that the $Z$ distribution has a small (if any) 
component at $Z \lsim 0.5$ \zsun (Casuso et al. 1996, Greggio 1997). The
SED in the UV offers another argument in support of this picture (see
also Bressan, Chiosi \& Fagotto 1994; Tantalo et al. 1996).

Turning now to the question of producing HHB stars at $Z \ge $ \zsun in
less than a Hubble time, an enhancement of the mass loss rate
parameter over the value which fits the GCs properties ($\eta_{\rm
GC}$) seems difficult to avoid. At low \dydz\ D'Cruz et al. (1996)
require an enhancement of a factor 2--3; at \dydz =2--3 Yi et al. 
require an enhancement of a factor $\sim 2$. It's interesting to
notice that, due to this large values of $\eta$, a small mass loss 
dispersion easily produces the hot helium flashers (see section
3). This class of objects naturally provides an upper limit to the
\te\ 
distribution on the HB (Castellani \& Castellani 1993), offering an
elegant solution to the problem of why the SED of giant ellipticals
shows the turnover at $\lambda \simeq 1000$ \AA. Indeed, if the HB were
populated down to the helium MS, stars would be distributed all the
way up to \te $\sim$ 50000 and beyond, hence producing a hard UV
spectrum shortward of Ly$\alpha$, which is not observed.

Thus, the hot stars in the nuclei of giant Es are likely to be objects in the
helium burning phase which happened to undergo a particularly heavy
mass loss while on the RGB. Their large $Z$ and $Y$ would produce
an evolution confined in a narrow range of effective temperatures. 
The hot edge of this range would be populated by objects which (due to mass
loss) failed helium ignition on the RGB, but succeeded later during
the evolution towards the WD stage. 
These stars would belong to
the high-$Z$ tail of the distribution in the GR90 picture, or,
alternatively, to the high mass loss tail of the 
distribution of stars around $\sim$
\zsun\ (DOR95). In the first case, the mass loss parameter $\eta$ should
increase with the metallicity; in the second a large dispersion of $\eta$
at Z$\sim$ \zsun is needed.
The first option more naturally accounts for the $(1550-V)$ vs \mg2
correlation, which we are going to consider next.

Finally, we attach a great significance to the fact that the bulk of
the UV emission in M31
bulge and M32 comes from objects which are fainter than the detection
threshold with FOC. Indeed, this leaves little alternative to HHB
stars as the main UV producers (cf. Fig. 1).

\subsection {On the $(UV-V)$ -- \mg2 correlation}

Among the various possibilities, the CSP in the nuclei of Es with 
\mg2$\gsim$ 0.3 can be modeled by a family of closed box
models, provided they are pre-enriched to $Z \sim$ \zsun\ (Greggio
1997). In these models, the metallicity distribution is $f(Z) \propto
exp(-Z/y)$, with $Z$ varying between a minimum value $Z_{\rm m}$
($\sim$ 0.5 \zsun), and a maximum value $Z_{\rm M}$. Here $y$ is the
yield as
defined by Tinsley (1980).  Since \mg2 is measured in the optical,
where low $Z$ stars have more weight, its value is very 
sensitive to $Z_{\rm m}$. The UV flux, instead, would be more
sensitive to $Z_{\rm M}$, if generated by stars in the high-$Z$ tail of the
distribution.  Therefore, the  $(UV-V)$ -- \mg2 correlation requires
that  $Z_{\rm m}$ and $Z_{\rm M}$ are well correlated, e.g. they both
increase with galaxy mass (luminosity), which seems plausible.
If HHB stars are produced only above a threshold metallicity at the
present age of the stellar populations in ellipticals, then 
the  $(UV-V)$ -- \mg2 correlation can result from the metallicity
distribution shifting to higher and higher values in galaxies with higher 
and higher \mg2 \ (GR90).

As for the galaxies with \mg2$\lsim 0.3$, there are very few
of them in the Burstein et al sample, and they define a correlation
with a different slope. From the \mg2 index one expects their
metallicity distribution to
be shifted to lower values, and thus it would be interesting to
know whether their UV
SED allows for a larger contribution from low $Z$ stars
(i.e. with intermediate temperatures, hence leading to flatter UV
upturns). To date this problem has not been quantitatively investigated. 

\subsection{The Evolution with Redshift Holds the Key}

The detection of the redshift evolution of the UV upturn remains
perhaps the most attractive opportunity for the future. By detecting
the effect we could in fact catch two birds with one stone. If indeed
the UV upturn fades away at $z\simeq 0.3\pm0.1$, this will represent
the decisive test for the HHB+AGB--manqu\'e origin of the upturn in
$z\simeq 0$ ellipticals. Moreover, the empirical
determination of the derivative $d(UV-V)/dz$ (hence of $d(UV-V)/dt$)
for galaxies of given value of
the central velocity dispersion $\sigma_\circ$,  could be used to set
constraints on the age dispersion among local ellipticals that would 
possibly be much tighter than those set by either optical colors or
the
fundamental plane relations.
The approach would be the same that Bower, Lucey \& Ellis (1992) have
pioneered to set such constraints using the small dispersion about the
average $(U-V)-\sigma_\circ$ relation of local cluster ellipticals,
with one advantage. Indeed, $U-V$ evolves very slowly in old
populations, i.e., by 0.02-0.03 mag/Gyr, while e.g. $(1550-V)$ should
evolve 10, perhaps 20 times faster. In principle, rest frame $UV-V$
colors
could set $\sim 20$ times tighter constraints to age dispersions.
However, the time derivative of 
$(1550-V)$ as determined from synthetic populations is extremely model
dependent, which therefore makes extremely attractive its direct,
empirical determination. We speculate that extensive studies of the UV
upturn for cluster vs. field ellipticals up to $z\sim 0.5$ could
greatly help tightening current constraints on the star formation
history of early-type galaxies.

\end{document}